\documentclass[%
preprint,
bibnotes,
 amsmath,amssymb,
 aps,
]{revtex4-2}

\usepackage{graphicx}
\usepackage{dcolumn}
\usepackage{epstopdf, epsfig}
\usepackage{amsmath,amsfonts,amssymb}
\usepackage{bm,url}
\usepackage[font=small,labelfont=bf,tableposition=top]{caption}

\newcommand\beq{\begin{equation}}
\newcommand\eeq{\end{equation}}
\newcommand\beqa{\begin{eqnarray}}
\newcommand\eeqa{\end{eqnarray}}
\newcommand{\ud}{\mathrm{d}}
\def\half {\textstyle{\frac{1}{2}}}

\newcommand{\irm}{\mathrm{i}}

\begin{document}

\title{Dynamically relevant recurrent flows obtained via a nonlinear recurrence function from two-dimensional turbulence.}

\author{Edward M. Redfern}
\affiliation{School of Mathematics and Computer Science, Keele University, Stafforshire, U.K. ST5 5BG}
\author{Andrei L. Lazer}
\author{Dan Lucas}
\affiliation{School of Mathematics and Statistics, University of St Andrews, Fife, U.K., KY16 9SS}\email{dl21@st-andrews.ac.uk}

\date{\today}

\begin{abstract}
This paper demonstrates the efficient extraction of unstable recurrent flows from two-dimensional turbulence by using nonlinear triads to diagnose recurrence in direct numerical simulations. Nearly recurrent episodes are identified from simulations and then converged using a standard Newton-GMRES-hookstep method, however with much greater diversity than previous studies which performed this `recurrent flow analysis'. Unstable periodic and relative periodic orbits are able to be identified which span larger values of dissipation rate, i.e. corresponding to extreme bursting events. The triad variables are found to provide a more natural way to weight the greater variety of spatial modes active in such orbits than a standard Euclidian norm of complex Fourier amplitudes. Moreover the triad variables build in a reduction of the continuous symmetry of the system which avoids the need to search over translations when obtaining relative periodic orbits. Armed with these orbits we investigate optimal weightings when reconstructing the statistics of turbulence and suggest that, in fact, a simple heuristic weighting based on the solution instability provides a very good prediction, provided enough dynamically relevant orbits are included in the expansion.
\end{abstract}

\maketitle
\section{Introduction}
One route to obtaining a simplified mathematical and physical description of fluid turbulence is by viewing a turbulent realisation as a trajectory through a high dimensional phase space \cite{Hopf:1948bn}. For sustained turbulence this trajectory is interpreted as moving between the neighbourhoods of simple invariant sets; unstable exact solutions to the equations of motion taking the form of equilibria, relative equilibria (travelling waves), time periodic orbits, relative periodic orbits or invariant torii. A chaotic trajectory is directed by the tangle of stable and unstable manifolds of such states and in this viewpoint the unstable exact coherent structures (ECSs) form the skeleton of a turbulent flow. 
We expect the turbulent trajectory to mirror the characteristics of ECSs during periods of close approach to them. Thus if ECSs can be isolated in sufficient abundance, they can form effective proxies for turbulent dynamics and can be used to make predictions about long-time statistics. 
Arguably the best candidates for such an analysis are recurrent flows, i.e. unstable periodic orbits (UPOs) and relative UPOs. These solutions exhibit dynamic cycles of behaviour (unlike fixed points) and can be computed by some shooting method to solve $X(t)-X(t-T)=0,$ where $T$ is the period of the orbit and $X$ some state vector.
Efforts to obtain such solutions in plane Couette flow \cite{Kawahara:2001ft, Viswanath:2007wc} and two-dimensional body-forced turbulence \cite{Chandler, Lucas:2015gt} demonstrated that it was possible for solutions to be representative of the turbulent state from whence they came. 

`Recurrent flow analysis' is thus the procedure of converging recurrent flows by first conducting a direct numerical simulation (DNS) and searching for nearly recurrent episodes to serve at starting guesses for a Newton based shooting iteration. This Newton method typically uses a Krylov method, e.g. GMRES, to solve the linear problem at each step as the system of equations cannot be formed explicitly since variations depend on the evolution of the Navier-Stokes system \cite{Viswanath:2007wc}. Additionally a hookstep is often applied to ensure each Newton step is within some trust region of the linearisation. In pioneering work, Chandler and Kerswell \cite{Chandler} utilised such a Newton-GMRES-hookstep (NGh) algorithm, combined with a search for near recurrences, to obtain relatively large numbers of recurrent flows from turbulent two-dimensional Kolmogorov flow. The recurrent flows were then used to reconstruct the statistics of the turbulent flow; for instance the probability density function (p.d.f.) of energetic quantities and mean flow profiles. Their conclusion was that, even at moderate Reynolds number, recurrent flows supporting high dissipation bursting events are missed, and a `democratic' equal weighting of the orbits in the reconstruction out-performed the cycle expansions of periodic orbit theory\cite{Cvitanovic:2013fz}.
A number of outstanding issues remained following the work of \cite{Chandler} and some follow-up investigations \cite{Lucas:2015gt, Lucas:2017fz}. 
The fact that `extreme' dynamics were not captured by recurrent flows is potentially rationalised by the hypothesis that such flows would be highly unstable and as such may not be shadowed closely enough for a full period in order for the recurrent flow analysis to identify an appropriate guess. The same reasoning holds for the inefficiency of the method as Reynolds number increases. However, if the current recurrence functions are missing near recurrences because of their choice of norm, this result may be a ``false-negative''. Therefore we wish to explore the possibility that the recurrence functions used previously are ill-suited for high dissipation events. Where a broad range of spatial Fourier modes are active, a simple $L_2$ norm of the difference between modes may not capture recurrence in small amplitude, but dynamically relevant modes.
In order to tackle some of these issues, recent research has been seeking alternative routes to isolate ECSs.  \cite{Farazmand:2016hf, Parker_Schneider_2022, Ashtari_Schneider_2023} employ variational methods to isolate recurrent flows; in the case of UPOs, loops in state space which do not satisfy the governing equations are iterated by adjoint-based minimisation of a cost function until the equations are satisfied. \cite{willis_2017, lucas_2022_stabilization, yasuda2024} have demonstrated that stabilisation of ECSs can be possible in some circumstances using feedback control.
The most notable advance was supplied by \cite{Page2024} who were able to identify a plethora of new recurrent flows in two-dimensional Kolmogorov flow which exhibit high dissipation and at high Reynolds number by making use of an automatic differentiation based gradient descent method as a preconditioner for guesses for NGh to effectively dispense with the need to identify near recurrences.

Along with the issues discussed above, relative periodic orbits (and travelling waves) can be isolated only when their translation is estimated when identifying near recurrence and when converging the exact solution. The `method of slices' \cite{Cvitanovic2012,Willis:2013bu} is a useful method developed, primarily for Galerkin-Fourier discretisations, to reduce continuous symmetries, and thereby avoid a search over translations. In developing new recurrence functions it would be advantageous to encode some symmetry reduction into the variables used. 

When studying recurrent flows \cite{Lucas:2015gt} demonstrated that cycles of behaviour can be dominated by key nonlinear interactions between Fourier modes in two-dimensional Kolmogorov flow. In particular, for a Galerkin-Fourier description of the Navier-Stokes equations we expect three-mode triad interactions due to quadratic nonlinearity \cite{Craik:2009p5}, i.e. in terms of wavevectors, mode $\bm{k}$ can exchange energy with $\bm{k}_1$ and $\bm{k}_2$ if $\bm{k} = \bm{k}_1 + \bm{k}_2.$ In \cite{Lucas:2015gt} the triad $(1,0)+(0,1)=(1,1)$ was found to be responsible for the majority of the nonlinear energy transfers in the recurrent flows identified, but with the mode $(1,1)$ playing a passive role. Such triad interactions were also discovered to predict bursting events in two-dimensional Kolmogorov flow by Farazmand \cite{Farazmand:2017ej} via the triad $(1,0)+(0,m) = (1,m).$ Dips in the energy of $(1,0)$ were found to be the precursor to high dissipation bursting events. 

The fact that the nonlinear transfers of energy between scales in a turbulent cascade are supported by triad interactions is a well-known and well-studied phenomenon \cite{Kraichnan_1959,Waleffe:1992dw,Moffatt_2014}. Much of the literature to-date has focused on the spectrum and fluxes of the Fourier amplitudes, however recent work has demonstrated that triad combinations of the Fourier phases play a key role in the dynamical behaviour, particularly extreme `bursting' events where synchronisation is observed \cite{Bustamante:2014jf, murray_2018, Protas_2024}. It is also known that, when interrogating the equations of motion, the key dynamical degrees of freedom are the so-called `triad phases'\cite{Bustamante:2009fm}; the Fourier phases do not evolve independently of each other but in three mode, triad combinations. The triad phases have the additional helpful property of filtering out any continuous symmetries of the system, the only difficulty is in determining an appropriate triad basis to use in any analysis \cite{Harper:2013gl}.

Given the above observations is seems reasonable that a recurrence function which encodes triad interactions may provide a useful alternative for identifying nearly recurrent flows in turbulent systems.

Once a sufficiently diverse and dynamically relevant set of recurrent flows have been identified, the next question is how to best use them to reconstruct the flow statistics. \cite{Cvitanovic:1992ih} lays out the framework of cycle expansions in `periodic orbit theory' as a means to accumulate the statistics of individual recurrent flows in a rigorous way. When this has been tested in fluid turbulence \cite{Chandler, Lucas:2015gt} it has not performed well, with the conclusion that insufficient orbits have been isolated for the expansion to converge. In these examples a uniform weighting was found to perform better than periodic orbit theory and other heuristic weightings based on the individual recurrent flow's linear instability. More recently, \cite{Page2024}, with a more comprehensive population of orbits, were able to show evidence for the Markovian view of a turbulent trajectory `pinball-ing' between invariant sets by computing a transition matrix from shadowing events and using its invariant measure as the weights of the recurrent flow expansion. While this is a motivational and significant result, in practical applications we may still desire a simplified model for the weights based on the intrinsic properties of the orbits. Furthermore, it has become clear that the overall number of orbits accumulated is not of primary importance, rather the orbits need to be sufficiently diverse to cover the turbulent attractor, including extreme events. The question which naturally arises is thus; what is the minimal set of recurrent flows that can be isolated and still well-represent the turbulent state?

The paper is organised as follows. Section \ref{sec:form} will introduce the equations of motion, diagnostics and key methods used. Section \ref{sec:recur_func} will present the formulation of a new `nonlinear` recurrence function based on triads and review its performance at isolating recurrent flows via recurrent flow analysis. In the course of this new high dissipation orbits are discovered and these are examined in detail in section \ref{sec:highD}. Section \ref{sec:stats} investigates some `optimal' ways to reconstruct statistics from recurrent flows, comparing weights chosen to minimise some loss functions and heuristic weights based on solution properties. We address the question of the minimal set of orbits and establish a clear connection between solution instability and preferred weights. A discussion section, \ref{sec:conclusion} concludes the paper.

\section{Formulation}\label{sec:form}

In this paper we will present the application of recurrent flow analysis to spatiotemporally chaotic Kolmogorov flow; the sinusoidally body forced incompressible two-dimensional Navier-Stokes equations. This flow is widely studied both for transition to turbulence and recurrent flow analysis. We consider a vorticity formulation for which the equations, in non-dimensional form, are

\begin{align}
\frac{\partial \omega}{\partial t} + \bm{u}\cdot\nabla \omega &= \frac{1}{Re} \Delta \omega -n \cos (n y) .\label{eq:NS}\\
\nabla\cdot \bm u &=0 \label{eq:incom}
\end{align}
with vorticity $\omega = \nabla \times \bm{u} \cdot \hat{\bm z}, $ velocity $\bm{u},$ $Re$ the Reynolds number. We will consider the periodic torus $[0,2\pi]\times[0,2\pi],$  a forcing wavenumber $n=4,$ and concentrate our efforts at the relatively well-studied $Re=40.$ The equations are solved with a standard pseudospectral method using two-thirds dealiasing, fourth order Runge-Kutta timestepping on the nonlinear and forcing terms and Crank-Nicolson on the viscous term. A resolution of $128^2$ is used and the code is implemented in CUDA to run on GPUs and available at \url{github.com/danl21/psgpu}. When converging invariant sets in the recurrent flow analysis, we use the NGh method developed in \cite{Chandler, Lucas:2015gt}. 

Throughout the course of the paper we will make use of the spectral representation of the fields of note, in particular we will note that

\begin{equation}
    \omega(\bm{x},t) = \sum_{\bm{k}} \Omega_{\bm k}(t) e^{i\bm{k}.\bm{x}}
\end{equation}
is the standard two-dimensional Fourier series for $\omega$ with wavevector $\bm{k}=(k_x,k_y)$ and $\Omega_{\bm k}$ are the complex Fourier coefficients which also obey $\Omega_{(-k_x,k_y)}=\Omega^*_{(k_x,k_y)}$ ($*$ denoting complex conjugation) to ensure the reality of $\omega.$ The number of active Fourier modes is set by the circular two-thirds dealiasing filter. 
The Kolmogorov flow system is invariant under the symmetries

\begin{align}\label{eq:sym}
\mathcal{S}:[u,v,\omega](x,y) &\rightarrow [-u,v,-\omega]\left(-x,y+\frac{\pi}{n}\right),\\
\mathcal{R}:[u,v,\omega](x,y) &\rightarrow [-u,-v,\omega]\left(-x,-y\right),\\
\mathcal{T}_s:[u,v,\omega](x,y) &\rightarrow [u,v,\omega]\left(x-s,y\right) \qquad \textrm{for } 0\leq s \leq {2\pi},
\end{align}

\noindent
where $\mathcal{S}$ represents the discrete shift-\&-reflect symmetry, $\mathcal{R}$ rotation through $\pi$ and $\mathcal{T}_s$ is the set of continuous translations by $s$ in $x$.

\subsection{Flow measures}

In order to analyse and compare the recurrent flows and turbulent statistics we define here some diagnostic quantities. Total energy, energy dissipation rate and energy input rate are defined in the standard way as
\begin{align}
E(t) &:= \half \langle \bm u^2 \rangle_V,\qquad
D(t) := \frac{1}{Re} \langle|\nabla \bm u |^2\rangle_V, \qquad
I(t) :=  \langle u \sin(ny) \rangle_V .
\end{align}
where the volume average is defined as 
$ \langle\quad\rangle_V := \frac{1}{4\pi^2} \int_0^{2\pi}\int_0^{2\pi} \ud x \ud y. $ Note the energy budget is such that $\ud E / \ud t = I-D$ meaning that any steady state, and the turbulent time average, must satisfy $D=I.$

For Kolmogorov flow the basic flow, which is the global attractor at small Reynolds number $Re,$ is given by the precise balance between forcing and dissipation; the profile and its energy and dissipation rate are 
\begin{align}
\bm u_{lam} := \frac{Re}{n^2}\sin ny \bm{\hat x}, \qquad \omega_{lam} := \frac{Re}{n}\cos ny \qquad E_{lam} := \frac{Re^2}{4n^4}, \qquad D_{lam} := \frac{Re}{2n^2}.
\label{eq:base}
\end{align}
For $n=4$, this flow becomes unstable at $Re\approx 10$ and a supercritical transition to turbulence is observed with $Re=40$ exhibiting sustained spatiotemporal chaos.

\section{Nonlinear recurrence function}\label{sec:recur_func}

Motivated by the desire for more efficient means to obtain recurrent flows, the requirement for accurate translations when obtaining relative periodic orbits and the idea that the typical variables used in numerical simulations may not be the most dynamically relevant, in this section we will explore definitions of alternative recurrence functions for observing near recurrences in direct numerical simulation. 
The recurrence function used throughout the majority of previous works which undertake the recurrent flow analysis makes use of an $L_2$ norm of the vorticity field. For instance, accounting for continuous translations in $x$ a suitable definition might read 
\begin{equation}\label{eq:Rog}
    R_\omega(t,\tau) = \min_{s\in [0,2\pi)} \frac{\sum_{\bm{k}} \left | \Omega_{\bm k}(t)e^{\irm k_x s} -\Omega_{\bm k}(t-\tau)\right|^2 }{\sum_{\bm{k}} |\Omega_{\bm{k}}(t)|^2 + \sum_{\bm{k}} |\Omega_{\bm{k}}(t-\tau)|^2}.
\end{equation}
Note, this has a slightly different denominator than that used in, e.g. \citep{Chandler, Lucas:2015gt}, chosen to ensure $0\leq R_\omega \leq 1.$ Variants can be defined where $\omega$ is interchanged with the relevant state vector (e.g. in \citep{LC} for three-dimensional stratified flows, the recurrence function requires a state vector composed of the three-dimensional vector vorticity and the buoyancy field).

In an effort to identify the truly dynamical degrees of freedom, and at the same time factor out the redundancy associated with translational invariance, we seek to utilise a recurrence function built from nonlinear triads. Examining the evolution equations for individual $\Omega_{\bm{k}};$

\begin{equation}
    \dot{\Omega}_{\bm{k}} = \frac{1}{2}\sum_{\bm{k}=\bm{k}_1+\bm{k}_2} \mathcal{Z}_{\bm{k}_1,\bm{k}_2}\Omega_{\bm{k}_1}\Omega_{\bm{k}_2} - \frac{|\bm{k}|^2}{Re}\Omega_{\bm{k}} + \frac{n}{2}\delta_{\bm{k}-(0,n)},
\end{equation}
where $\mathcal{Z}_{\bm{k}_1,\bm{k}_2} = \left (\frac{1}{|\bm{k}_1|^2} - \frac{1}{|\bm{k}_2|^2} \right)\bm{k}_1 \times \bm{k}_2$ and $\delta_{\bm{k}}$ is the Kronecker delta function. The sum is over triads using the standard convention over repeated indices. Considering the amplitude and phase of each complex Fourier mode, $\Omega_{\bm{k}}(t)=a_{\bm{k}}(t)\mathrm{e}^{\irm \phi_{\bm{k}}(t)},$ we can write 

\begin{align}
    \dot{a}_{\bm{k}} &=  \frac{1}{2}\sum_{\bm{k}=\bm{k}_1+\bm{k}_2} \mathcal{Z}_{\bm{k}_1,\bm{k}_2}a_{\bm{k}_1}a_{\bm{k}_2}\cos(\varphi_{\bm{k}_1,\bm{k}_2}^{\bm{k}}) - \frac{|\bm{k}|^2}{Re}a_{\bm{k}} -2 \cos(\phi_{\bm{k}})\delta_{\bm{k}-(0,n)}, \label{eq:amp}\\
    \dot{\phi_{\bm{k}}} &= \frac{1}{2}\sum_{\bm{k}=\bm{k}_1+\bm{k}_2} \mathcal{Z}_{\bm{k}_1,\bm{k}_2}\frac{a_{\bm{k}_1}a_{\bm{k}_2}}{{a}_{\bm{k}}}\sin(\varphi_{\bm{k}_1,\bm{k}_2}^{\bm{k}}) +\frac{2}{a_{\bm{k}}} \sin(\phi_{\bm{k}})\delta_{\bm{k}-(0,n)}, \label{eq:phase}
\end{align}
where $\varphi_{\bm{k}_1,\bm{k}_2}^{\bm{k}} = \phi_{\bm{k}_1} + \phi_{\bm{k}_2} - \phi_{\bm{k}}$ is the triad phase for the triad $\bm{k}=\bm{k}_1 + \bm{k}_2.$ It can be shown that by combining equations \eqref{eq:phase}, a closed system of equations can be written for $\varphi_{\bm{k}_1,\bm{k}_2}^{\bm{k}}$ indicating that the key degrees of freedom are not the individual Fourier phases but the triad phases \cite{Bustamante:2009fm}. If we begin with $N$ total Fourier modes and seek to construct a linearly independent triad basis, we will find that there will be $N-1$ linearly independent triad phases. The degree of freedom which has been reduced is the one associated with the translational invariance in $x$ meaning that the real amplitudes $a_{\bm{k}}$ and triad phases $\varphi_{\bm{k}_1,\bm{k}_2}^{\bm{k}}$ form a natural symmetry reduction. To observe the symmetry reduction in the amplitude/phase representation, a translation in $x$ by a shift $s$ is a rotation of the complex Fourier amplitude of mode $(k_x,k_y)$ by an angle $k_xs,$ i.e. 

\begin{align}
    \varphi_{\bm{k}_1,\bm{k}_2}^{\bm{k}} \to& (\phi_{\bm{k}_1} + k_{1,x}s) + (\phi_{\bm{k}_2} + k_{2,x}s) - (\phi_{\bm{k}} + k_{x}s)\\ &= (\phi_{\bm{k}_1}  + \phi_{\bm{k}_2}- \phi_{\bm{k}}) + (k_{1,x}+ k_{2,x}- k_{x})s\\ &= \varphi_{\bm{k}_1,\bm{k}_2}^{\bm{k}} ,
\end{align}
 since $k_{1,x}+k_{2,x}-k_x=0$. Note that, in this system, the flow is pinned in the $y$ direction by the forcing which breaks the continuous symmetry of the Navier-Stokes equations in $y$; had this not been the case the number of triad phases would be $N-2$ accounting for invariance in $x$ and $y.$ 

 Lucas and Kerswell\citep{Lucas:2015gt} determined that recurrence functions constructed from real amplitudes, $a_{\bm{k}}$ only, are not effective for diagnosing recurrence. We expect only using triad phases to be even less effective; their rapid evolution introduces considerable noise into any recurrence measure which uses them directly. The difference in dimension/scale of the amplitudes and phases means that care needs to be taken in creating a measure which might combine these quantities. The obvious combination to use is a triad amplitude

 \begin{align}
     \xi_{\bm{k}_1,\bm{k}_2}^{\bm{k}} = \Omega_{\bm{k}_1}\Omega_{\bm{k}_2}\Omega^*_{\bm{k}} = a_{\bm{k}_1}a_{\bm{k}_2}a_{\bm{k}} \mathrm{e}^{\irm \varphi_{\bm{k}_1,\bm{k}_2}^{\bm{k}}}
 \end{align}
which gives the appropriate amplitude weighting for a given three-mode triad while retaining the correct triad phase combinations. Note this product arises when examining energy flux between Fourier modes \cite{Buzzicotti:2016bz,murray_2018}. Therefore our recurrence function based on Fourier triads can be defined as

\begin{equation}
    R_{\xi}(t,\tau) = \frac{\sum_{\mathcal{P}} \left| \xi_{\bm{k}_1,\bm{k}_2}^{\bm{k}} (t) -\xi_{\bm{k}_1,\bm{k}_2}^{\bm{k}} (t-\tau)\right|^2 }{\sum_{\mathcal{P}} |\xi_{\bm{k}_1,\bm{k}_2}^{\bm{k}}(t)|^2 + \sum_{\mathcal{P}} |\xi_{\bm{k}_1,\bm{k}_2}^{\bm{k}}(t-\tau)|^2},
\end{equation}
where the sums are over an appropriately obtained set of linearly independent triads, $\mathcal{P}$. Note that no special treatment of symmetries is required; $R_\xi$ is invariant to both continuous translations in $x$ and even multiples of shift-reflect $\mathcal{S}.$ As has been considered previously, odd multiples of shift-reflected orbits will be preperiodic to an even multiple version of extended period and so are not searched for.

In order to practically construct $R_\xi$ we first need to compute an appropriate set of triads $\mathcal{P}$. The method we use follows \cite{Harper:2013gl,MurrayPhD} in computing a linear system of triad equations $k_1+k_2-k_3=0$ and performing a rank reduction to compute the null space using a QR decomposition of the so-called `cluster matrix'. This can be a costly exercise, and although the problem is very sparse and only needs to be carried out once ever for any given resolution, we limit the size of the set to 655 triads, instead of a full basis of  2707 (the count after circular 2/3 dealiasing). The 655 triads are those formed using only the first half of the wavenumbers in each direction, i.e. $\mathcal{P}$ forms a full basis for a $64^2$ resolution. We essentially decide to neglect the very large $k$ as they will be very low amplitude and dominated by viscosity (more discussion on this in section \ref{sec:highD}). The code for computing the triad set, and the triad set itself, is included in supplemental material. 

By way of fair comparison we will also define the symmetry reduced recurrence function using the method of slices. This will use the first Fourier mode slice, i.e. defining $\bar\phi(t) = \phi_{(1,0)}(t)$ the recurrence function \ref{eq:Rog} is modified such that

\begin{equation}
    R_{\phi}(t,\tau) = \frac{\sum_{\bm{k}} \left| \Omega_{\bm k}(t)e^{-\irm \bar{\phi}(t) } -\Omega_{\bm k}(t-\tau)e^{-\irm \bar{\phi}(t-\tau) }\right|^2 }{\sum_{\bm{k}} |\Omega_{\bm{k}}(t)|^2 + \sum_{\bm{k}} |\Omega_{\bm{k}}(t-\tau)|^2}.
\end{equation}
This ensures that the Fourier modes lie on the slice where $\bar\phi=0$ when computing the recurrence function, thereby reducing translations in $x.$ 

Near recurrences are then generated in the standard way using $R_\xi$ and $R_\phi,$ by setting a threshold and searching back through the DNS history (of 50 time units). The two cases are performed on the same turbulent trajectory, i.e. beginning from the same initial condition.  Minima over $\tau$ are obtained from episodes when $R$ is below this threshold. These are output for processing by Newton-GMRES-hookstep. Note NGh operates on the $\Omega_{\bm{k}}$ variables with shift $s$ as part of the system of unknowns, in-keeping with \citep{Chandler} etc. In order to obtain an estimate of the starting shift from the symmetry reduced recurrences we use an averaging of Fourier phase rotations as described in \cite{lucas_2022_stabilization}. Discrete shift-reflects are found by checking the minimum over the $n$ possible $y$ shifts.

\begin{figure}
    \centering
    \includegraphics[width=0.45\linewidth]{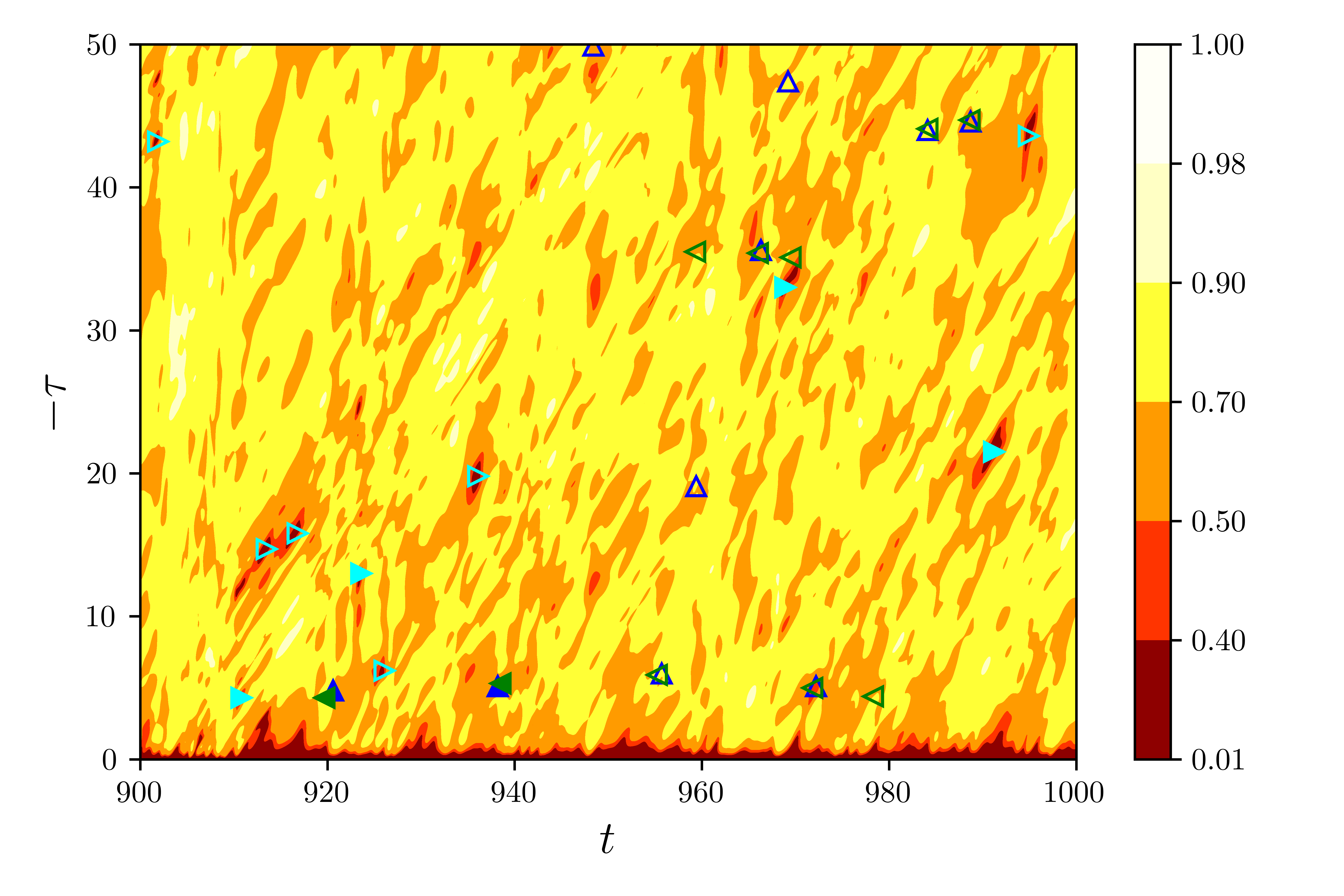}
    \includegraphics[width=0.45\linewidth]{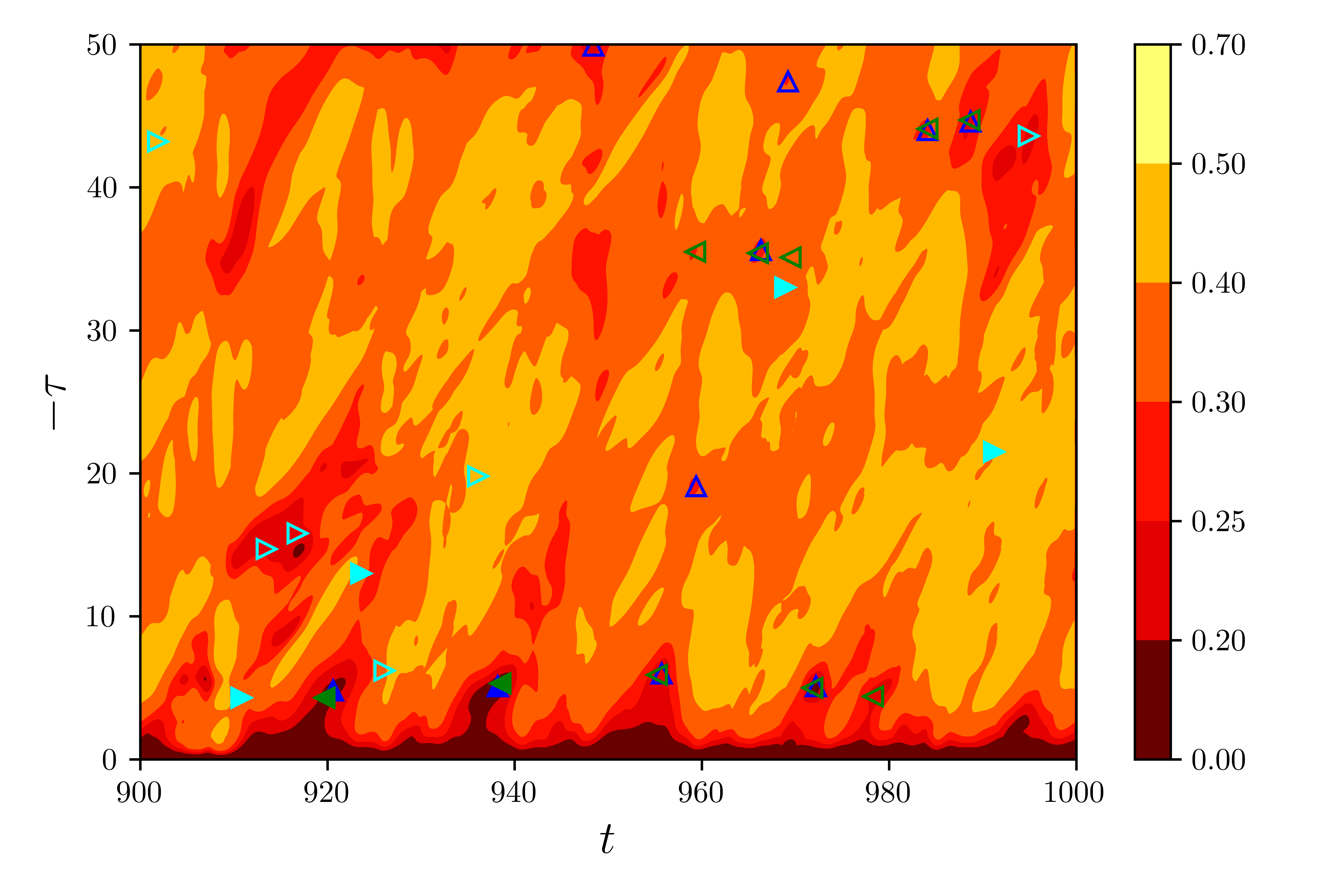}
    \caption{Plots of $R_\xi(t,\tau)$ (left) and $R_\phi(t,\tau)$ (right). Identified near recurrences are indicated by; circles from $R_\phi$, triangles from $R_\xi$. Filled symbols are near recurrences that are successfully converged using NGh.}
    \label{fig:R_t_tau}
\end{figure}

Figure \ref{fig:R_t_tau} shows plots of $R_\xi$ and $R_\phi$ for a certain short window where near recurrences are identified and then attempted to be converged. Filled symbols show successful convergence, open unsuccessful. Triangles represent those identified using $R_\xi$ and circles using $R_\phi.$ It is clear that $R_\xi$ effectively identifies several episodes of near recurrence, shadowing unstable periodic orbits, which are missed by $R_\phi$ (or $R_\omega,$ not shown). The minima of $R_\xi$ appear to be more distinct and deeper than those of $R_\phi.$ 

We perform a  DNS of 25 000 time units and, by tuning the threshold for each recurrence function, attempt to obtain a similar number of nearly recurrent episodes for NGh to attempt to converge. Setting a threshold $R_\xi<0.4$ for the triad recurrence function yields 1654 guesses for nearly recurrent flows and $R_\phi<0.3$ using the symmetry reduced recurrence function using the method of slices yields 1358 guesses. On processing with NGh, there is a consistent success rate, around $\frac{1}{4}$ of guesses are converged, with 47 unique solutions generated from $R_\xi$ recurrences and 43 by $R_\phi$. The triad recurrence function does not seem to provide better conditioned guesses for NGh overall, at least at this threshold. However, there are differences in the solutions which are obtained. 

\begin{figure}
    \centering
    \includegraphics[width=0.45\linewidth]{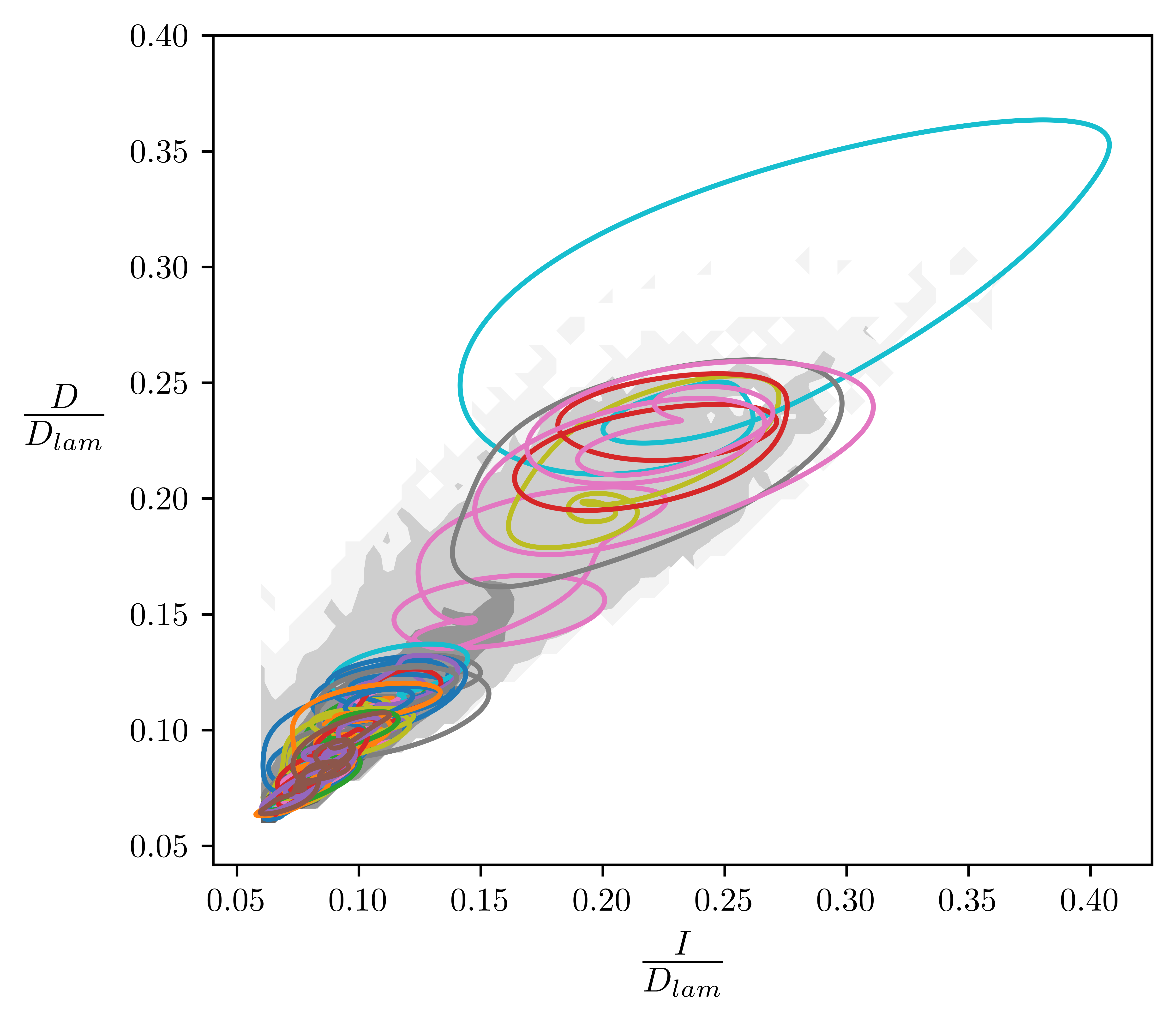}
    \includegraphics[width=0.45\linewidth]{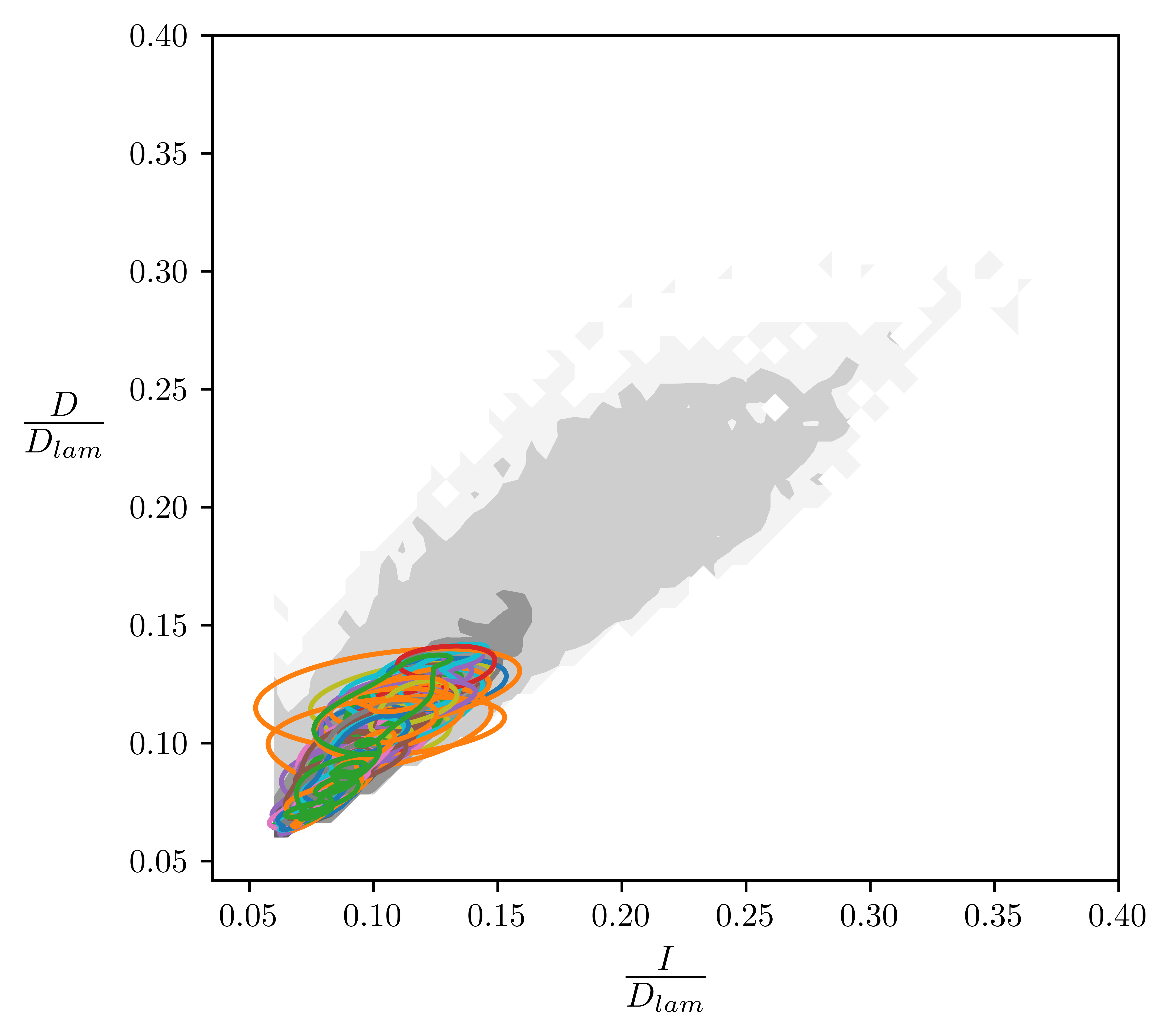}
    \caption{Projection of recurrent flows onto the $(D/D_{lam}, I/I_{lam})$ plane, left from $R_\xi$ and right from $R_\phi$. These are respectively the 47 and 43 solutions obtained from the DNS of 25, 000 time units described in the text.}
    \label{fig:DI_UPOs}
\end{figure}

\begin{figure}
    \centering
    \includegraphics[width=0.95\linewidth]{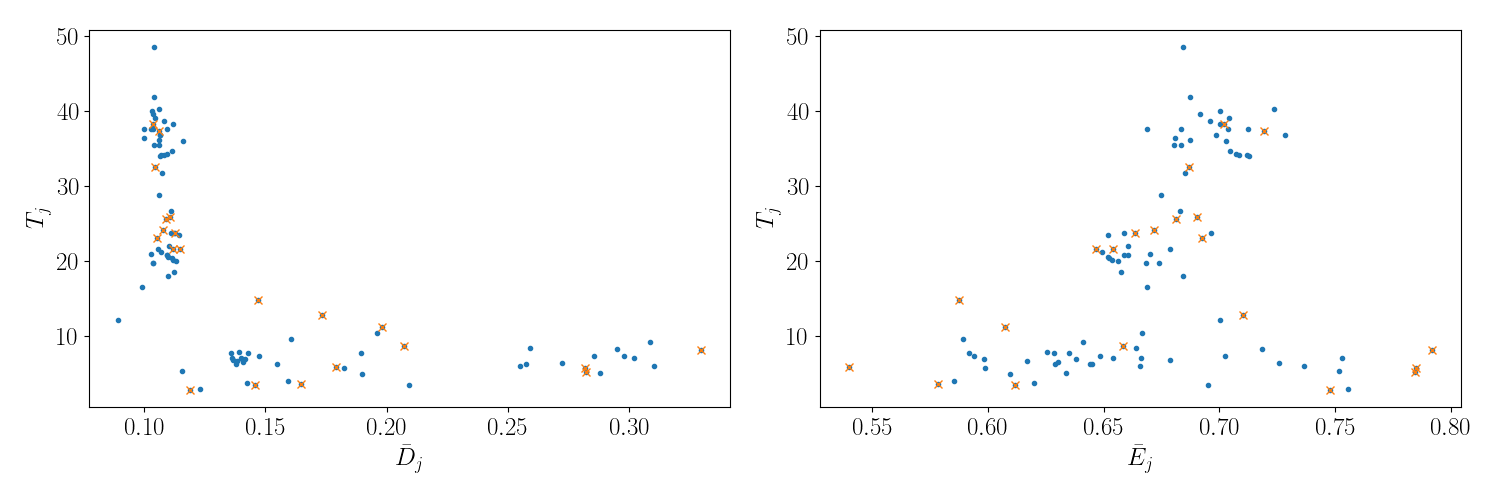}
    \caption{Plots showing recurrent flow period against mean dissipation rate $\bar{D}$ (left) and mean total energy $\bar{E}$ (right). Extreme states have lower period, and longer periods are dominated by ordered states (lower $\bar{D}$ and intermediate $\bar{E}$).}
    \label{fig:Period}
\end{figure}

A striking visualisation of this is shown in figure \ref{fig:DI_UPOs} where we plot all recurrent flows (UPOs and relative UPOs) projected onto the plane $(I/D_{lam}, D/I_{lam}),$ left plot showing those states converged from $R_\xi$ near recurrences and right from $R_\phi, $ with the p.d.f. of the turbulent attractor plotted in grey in the background. A well-reported issue in previous work performing recurrent flow analysis in Kolmogorov flow is a lack of orbits which exhibit high dissipation and extreme events \citep{Chandler, Lucas:2015gt}. Figure \ref{fig:DI_UPOs} shows that the solutions obtained from $R_\xi$ provide significantly better coverage of the high dissipation region compared to those obtained using $R_\phi.$ Orbits obtained using $R_\phi$ occupy only the region $D/D_{lam}<0.15$ whereas five of those obtained using $R_\xi$ reach up to $D/D_{lam}\approx 0.25$ and above. These high dissipation solutions represent bursting events and are, unsurprisingly, significantly more unstable than the low dissipation solutions (see suplemental material for the full catalogue and properties). Consistently with \cite{Page2024} we find high dissipation orbits have relatively low periods ($<10$), and a clear clustering is observed when plotting period against mean energetics; extremal orbits are always of short period and long periods are always restricted to regular dynamics, see figure \ref{fig:Period}.

Motivated by this success we go on to perform a longer DNS and produce a larger set of guesses for NGh using $R_\xi$ (only). Running for 60 000 time units with the same parameters as before, with a threshold $R_\xi<0.4$ to identify a period of near recurrence. This yields 4306 near recurrences from which 101 distinct recurrent flows are obtained (plus many which converge to known equilibrium and travelling wave solutions). For context \cite{Chandler}, at the same parameters, from a DNS of $10^5$ time units, were able to obtain 58 unique recurrent flows. Moreover the success of obtaining a greater diversity of solution is continued with 14 solutions having high mean dissipation. 
\subsection{High dissipation orbits}\label{sec:highD}

\begin{figure}
    \centering
    \includegraphics[width=0.95\linewidth]{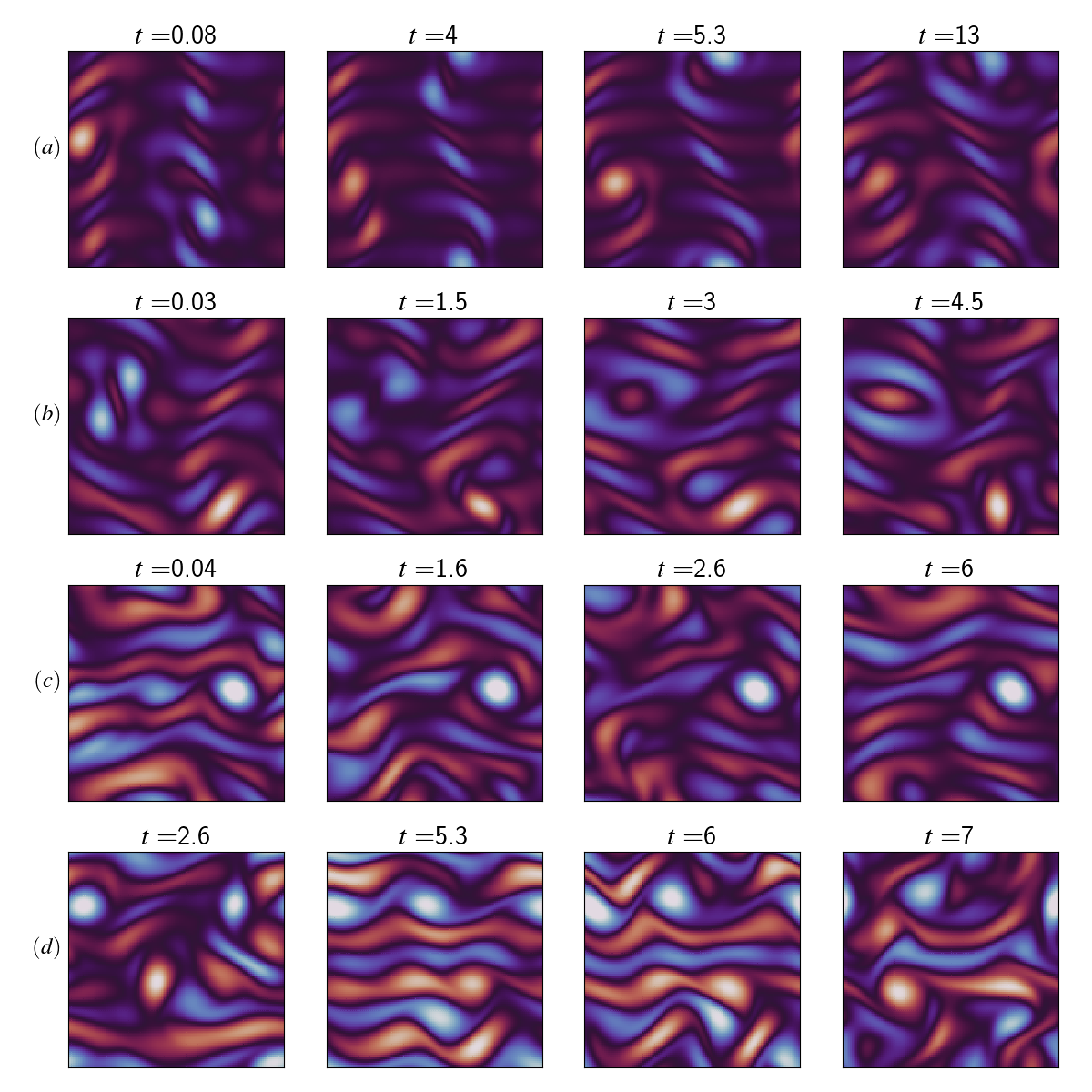}
    \caption{Snapshots of vorticity of four representative recurrent flows, moving from low dissipation in the top row to the largest at the bottom. The four recurrent flows are plotted in figure \ref{fig:DIsnapshots} with symbols to indicate the time of each snapshot. Orbits have periods 16.56, 6.94, 8.74, 8.20 from top to bottom. Colour limits $[-10,10]$ for all plots.}
    \label{fig:snapshots}
\end{figure}

\begin{figure}
    \centering
    \includegraphics[width=0.95\linewidth]{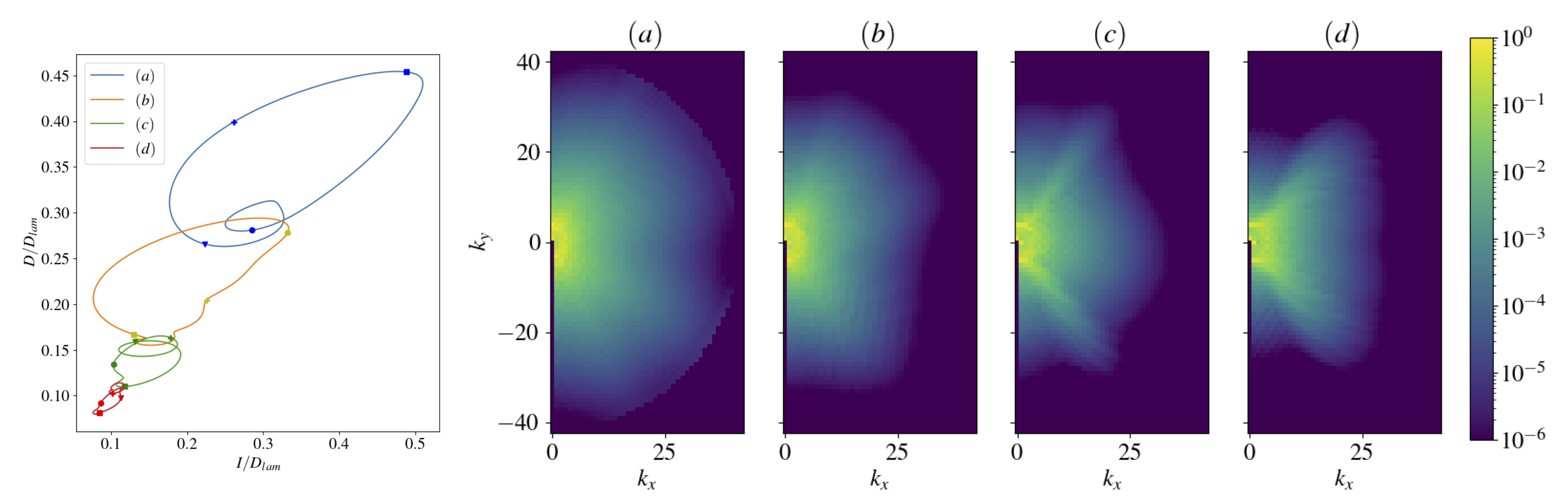}
\caption{Left: Dissipation rate against energy input rate for the four recurrent flows shown in figure \ref{fig:snapshots}. Symbols represent the columns/times in figure \ref{fig:snapshots}, from earliest to latest: circle, triangle, cross, square. Right: Two-dimensional spectra of the variance of Fourier amplitudes, $V(a_k)$ for the four orbits (a)-(d).}
    \label{fig:DIsnapshots}
\end{figure}

In order to compare and contrast the higher dissipation orbits to the more regular ones, and to attempt to establish why $R_\xi$ is more effective at picking them out, we identify four representative orbits which span from low to high dissipation rate, which we label (a), (b), (c) and (d), (a) being the recurrent flow that attains the largest dissipation rate and (d) a generic low dissipation solution. Figure \ref{fig:DIsnapshots} shows the projection onto the $(D/D_{lam}, I/I_{lam})$ plane of these four orbits which have periods respectively  16.56, 6.94, 8.74, 8.20. We then plot four snapshots of vorticity in figure \ref{fig:snapshots} for each orbit. A clear progression is observed moving from a relatively low amplitude state to a much more energetic state with several isolated vortices riding on jet-like streamwise flow structures induced by the forcing. The lower amplitude states are similar to those observered in \cite{Chandler}, with a spatial structure similar to the kink-antikink states of \cite{Lucas:2014ew}, and are expected to be dominated by the nonlinear triad interaction $(0,1)+(1,0)=(1,1)$ \cite{Lucas:2015gt,Page2021}. The higher amplitude states show more variety across a period with more obvious isolated vortices as $\bar{D}$ increases combined with a clear mode 4 jet pattern due to the forcing. 


In the more disordered, high dissipation bursting flows, a greater variety of modes are active in the periodic cycle of high dissipation. To demonstrate this we define the variance of the individual Fourier amplitudes

$$
V(a_k) = \frac{1}{T}\int_0^T(a_k-\bar{a}_k)^2 \ud t,
$$
where $\bar{a}_k = \frac{1}{T}\int_0^T a_k \ud t$ and plot the two-dimensional spectra for the four orbits. Unsurprisingly (a) demonstrates a broader band and more isotropic spectrum of active modes, with (d) much narrower, less isotropic ($k_x=k_y$ modes are more active) and having a smaller number of small wavenumbers which are active. Close inspection of (d) confirms high variance for modes (0,1) and (1,0), consistent with the findings of \cite{Lucas:2015gt}, in addition to the forced mode (0,4) and its harmonics. The larger $D$ orbits have a wider selection of small $k$ modes active, suggesting that a wider variety of nonlinear interactions are taking place during the cycle of dynamics, as we may expect from the vorticity snapshots in figure \ref{fig:snapshots}. 

It is therefore crucial that a recurrence function can accurately identify near recurrence across this range of modes and an $L_2$ norm of complex Fourier coefficients will naturally down-weight smaller amplitude but dynamically important modes. Accumulating these modes via their dominant interaction combinations can balance modes of differing energy/enstrophy in a dynamically meaningful way, and also avoid the impact of, and sensitivity due to, the faster evolution of phases relative to amplitudes. 

\section{Recreating turbulent statistics from recurrent flows}\label{sec:stats}

\begin{figure}
    \centering
    \includegraphics[width=0.95\linewidth]{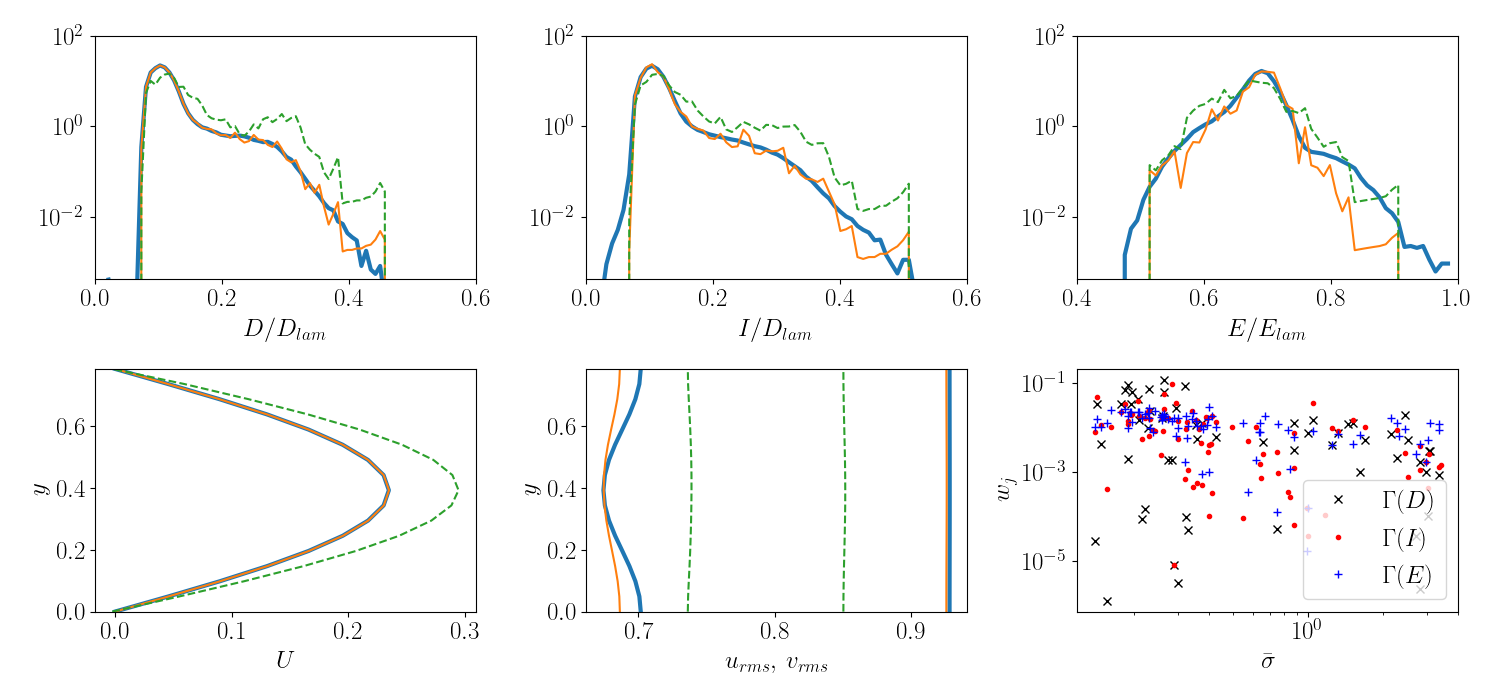}
    \caption{Predictions of flow statistics using recurrent flows. Top row, p.d.f.s of dissipation rate $D,$ energy input rate $I$ and energy $E$, bottom row symmeterised mean flow $U(y)$ and r.m.s velocity components $u_{rms}$ and $v_{rms}.$ Thick blue line is the ground truth from DNS, orange is the KL-divergence optimal and green dashed is the control of uniform weighting $w_j=1/N.$ This case uses all recurrent flows and minimises $\mathcal{L}(D).$ Bottom right shows the computed weights against the sum of the real parts of unstable Floquet exponents for the cases minimising $\mathcal{L}(D),$ $\mathcal{L}(I),$ and $\mathcal{L}(E).$ }
    \label{fig:full_pred}
\end{figure}

Having obtained a high quality set of recurrent flows we seek to establish their ability to predict the statistics of the turbulent attractor. For a given statistic $\Gamma$ we seek a prediction based on an expansion in terms of $N$ recurrent flows

\begin{equation}
    \Gamma_{pred} = \sum_{j=1}^N w_j \Gamma_j
\end{equation}
where the weights $w_j$ are such that $\sum_{j=1}^N w_j=1$ and $\Gamma_j$ is the statistic obtained for each individual solution. In previous work \cite{Chandler, Lucas:2015gt} an even weighting of recurrent flows (i.e. $w_j=1/N$) was found to outperform cycle expansions of periodic orbit theory when reconstructing probability density functions of energetic quantities. More recently \cite{Page2024} found that using weights derived from the invariant measure of transition probabilities of shadowing episodes of the recurrent flows was able to produce a set of weights that effectively reproduce the statistics. However, while this demonstrates the Markovian view of turbulence moving between neighbourhoods of recurrent flows, in practice it would be preferable to have a simpler model for weighting UPOs based on their intrinsic properties. To this end, we will investigate some ``optimal'' expansions (following \cite{page2023}) by obtaining the weights which minimise the Kullback-Leibler divergence (or KL-divergence)

\begin{equation}
    \mathcal{L}(D) = \int \Gamma(D)\log \left ( \frac{\Gamma_{pred}(D)}{\Gamma(D)} \right ) + \Gamma(D) - \Gamma_{pred}(D) \, \ud D,
\end{equation}
which is a loss function designed for minimising the distance between probability distributions \cite{Boyd_Vandenberghe_2004}. We use the scipy minimiser \cite{2020SciPy-NMeth} with the L-BFGS-B algorithm\cite{Liu_1989}. In this context $\Gamma(D)$ will be the probability density function for the dissipation rate $D,$ but other p.d.f.s, e.g. of $E$ or $I$, can and will be trialled. 
Figure \ref{fig:full_pred} shows the results of this procedure, arguably showing the best agreement to date at this Reynolds number. Note here we find weights minimising $\mathcal{L}(D),$ and use those weights to reconstruct the other statistics, therefore, while the p.d.f. of $D$ shows excellent agreement the other statistics are also very well approximated, certainly much better than the control weighting, $w_j=1/N.$ 

Also shown in figure \ref{fig:full_pred} are the weights $w_j$ plotted against the sum of the real parts of the unstable Floquet exponents for each recurrent flow $j;$ 

\begin{equation}
    \bar{\sigma} = \sum_i \sigma_i,
\end{equation}
where $\sigma_i$ are the real parts of individual unstable exponents.
Cases minimising $\mathcal{L}(D),$ $\mathcal{L}(I),$ and $\mathcal{L}(E)$ are all shown on the same axes. A weak negative correlation is observed, as expected the more unstable, extremal orbits are less strongly weighted, but considerable scatter is seen in the data. In particular the optimal set has many orbits with very small weights. On closer inspection the minimisation algorithm has returned a significant number of weights which are identically zero. For $\mathcal{L}(D)$ 47 orbits are given zero weight,  $\mathcal{L}(I),$ 13 and $\mathcal{L}(E)$ 14. It is somewhat remarkable that the prediction is as accurate as it is with only 49 out of 101 flows having $w_j>10^{-4}.$

By way of quantifying the quality of the fit, independently of the KL-divergence, we define the mean squared error

\begin{equation}
    \mathcal{E}(\Gamma) = \frac{1}{n}\sum_i^n (\Gamma - \Gamma_{pred})^2
\end{equation}


where $n$ are the number of records in the statistic $\Gamma$ (e.g. bins of the p.d.f or grid points in the mean profile). Table \ref{tab:tab1} records $\mathcal{E}$ for various statistics and the three loss functions considered, plus the uniform weighting. 

These observations prompt the following questions; what is the smallest number of orbits required to approximate the statistics to a given accuracy? And, since the minimiser is not guaranteeing a global optimal, can reducing the number of orbits in the expansion give a clearer indication of correlations with e.g. flow instability?

\begin{figure}
    \centering
    \includegraphics[width=0.95\linewidth]{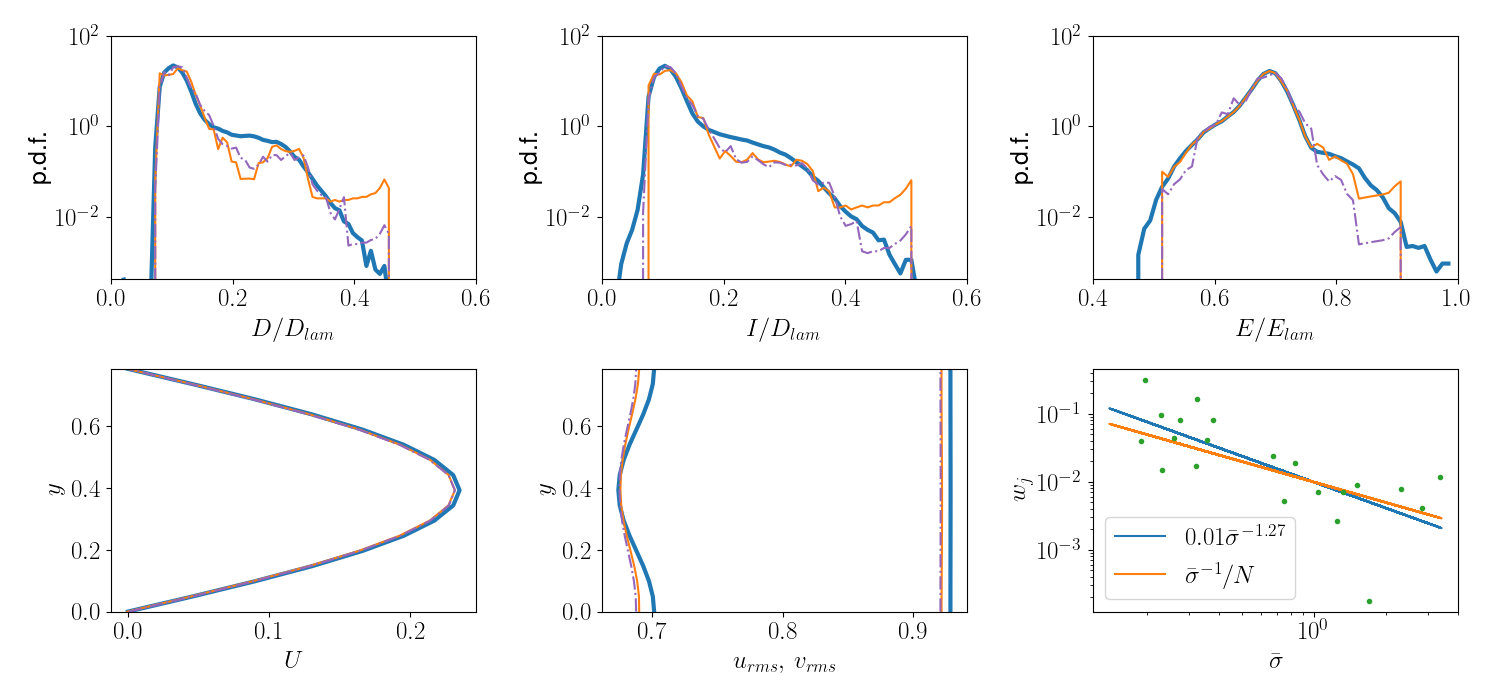}
    \caption{Predictions of flow statistics using recurrent flows. Top row, p.d.f.s of dissipation rate $D,$ energy input rate $I$ and energy $E$, bottom row symmeterised mean flow $U(y)$ and r.m.s velocity components $u_{rms}$ and $v_{rms}.$ Thick blue line is the ground truth from DNS, orange is the KL-divergence optimal for $\mathcal{L}(E)$ using the pruned set of 20 most relevant recurrent flows and purple dot-dashed is the expansion using $w_j=\bar{\sigma}_j^{-1}/\sum_j^N\bar{\sigma}_j^{-1}$ and all 101 recurrent flows found. Bottom right shows the computed weights against $\bar{\sigma}$ with a line of best fit and the line $0.01\bar{\sigma}_j^{-1}$. }
    \label{fig:prune_pred}
\end{figure}

\begin{figure}
    \centering
    \includegraphics[width=0.75\linewidth]{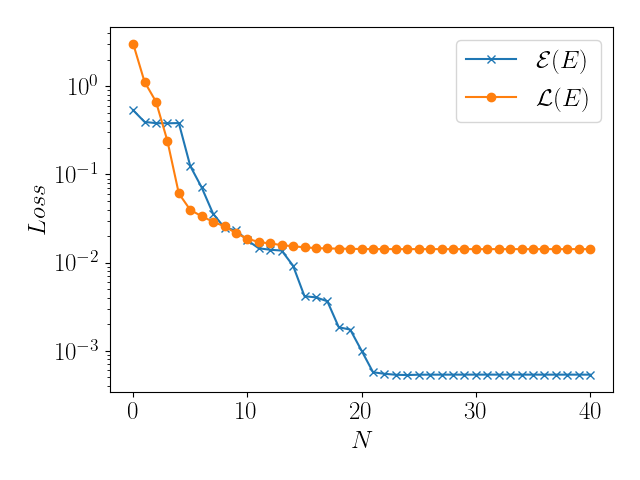}
    \caption{plot showing the KL-divergence loss and the mean squared error, both on energy $E,$ against the number of recurrent flows as the pruning strategy is performed.}
    \label{fig:prune}
\end{figure}

To address this we investigate a pruning strategy to find a minimal set of orbits which can reconstruct the statistics effectively. Using $E$ results in relatively few vanishing weights, and it is arguably more difficult to reconstruct with UPOs compared to $D$  therefore we switch to using $\mathcal{L}(E)$ as our loss function. Our procedure is to initialise a list with the pair of orbits which minimises $\mathcal{L}(E)$ and then iteratively append the orbit which reduces the loss the most, across all remaining solutions. Figure \ref{fig:prune} shows the KL divergence and mean squared error, $\mathcal{E}(E),$ for this procedure as function of the number of orbits in the expansion. We find that 20 orbits are sufficient to provide an accurate representation of the energy p.d.f., both measures fail to show significant improvement beyond this point. Figure \ref{fig:prune_pred} shows the reconstruction of the statistics using these 20 orbits, demonstrating surprisingly good performance. It should be mentioned that, as indicated by figure \ref{fig:prune}, reasonable approximations are possible with fewer than 20 orbits. Table \ref{tab:tab1} shows that, in terms of the mean squared error, this expansion is as good, if not better, than most of the examples using many more solutions. Moreover, figure \ref{fig:prune_pred} also shows a plot of $w_j$ against $\bar{\sigma}_j,$ with this reduced set. We now observe a much clearer negative correlation and a line of best fit indicates a trend $~\bar{\sigma}^{-1}$ (the best fit exponent is -1.27, but ignoring the smallest weight gives -1.06). As a result, we have also reconstructed the expansion using $w_j \propto \bar{\sigma}_j^{-1}$ shown in purple in figure \ref{fig:prune_pred} and in table \ref{tab:tab1}, now with all 101 recurrent flows obtained. The performance of this approximation is, again, very good. Such an `escape-time' weighting was suggested by \cite{Zoldi_1998} and used in \cite{Chandler} but dismissed as being no better than a uniform weighting. It is clear from these results that with the right orbits a weighting inversely proportional to $\bar{\sigma}$ is at least as good as the minima of $\mathcal{L}.$ 
We should note at this point that other correlations were checked in the data. For instance \cite{Kazantsev:1998wb, Chandler} suggest $w_j\propto T \bar{\sigma}_j^{-1},$ to weight longer period orbits more strongly. One might imagine that the Floquet multiplier, $\exp(\sigma T)$, or its inverse, could also provide a meaningful heurstic weighting. However none of these were found to perform as well as $\bar{\sigma}^{-1}.$ Likewise the cycle-expansions of periodic orbit theory were trialled using this data set and showed less successful reconstructions than the ones shown here, severely under estimating the contributions from the extreme events.
\begin{table}
    \begin{tabular}{l|ccccccc}
      $w$ & $\mathcal{E}(D)$ & $\mathcal{E}(I)$ & $\mathcal{E}(E)$ & $\mathcal{E}(U)$ & $\mathcal{E}(u_{rms})$ & $\mathcal{E}(v_{rms})$ & $w\neq 0$\\
      \hline
        $\mathcal{L}(D)$ & 0.0012 & 0.12 & 0.51 & 2.1e-8 & 9.8e-5 & 5.5e-6 & 54\\
        $\mathcal{L}(I)$ & 0.29 & 5.1e-4 & 0.19 & 5.8e-8 & 7.1e-5 & 3.3e-5 & 88\\
        $\mathcal{L}(E)$ & 0.69 & 0.22 & 5.3e-4 & 1.7e-4 & 1.2e-4 & 4.9e-4 & 87\\
        $1/N$ & 3.7 & 2.2 & 1.8 & 0.0017 & 0.0025 & 0.006 & 101 \\
        $\mathcal{L}^{prune}(E)$ & 2.1 & 0.85 & 1.7e-3 & 5.4e-6 & 5.3e-5 & 4.5e-5 & 20\\
        $\bar{\sigma}^{-1}$ & 0.75 & 0.27 & 0.28 & 5.2e-6 & 7.8e-5 & 5.8e-5 & 101\\
    \end{tabular}
    \caption{Table showing mean squared errors, $\mathcal{E},$ for various reconstructions of the statistics. $w$ column indicates which method is used, e.g. $\mathcal{L}$ being the loss function optimised. The column $w\neq0$ indicates how many recurrent flows are active in the reconstruction, e.g. the pruned case only use 20 and those not constructed from a minimisation use all 101.}
    \label{tab:tab1}
\end{table}

\section{Conclusions and discussion}\label{sec:conclusion}
In this paper we have shown that, by judicious choice of variable, recurrent flow analysis can be significantly improved. In particular, high dissipation, rare, bursting recurrent flows can be identified by near recurrence, something missing in previous work \cite{Chandler, Lucas:2015gt} and only recently obtained in \cite{Page2024} by a gradient descent preconditioning of guesses. 
By interrogating the variance of Fourier modes, we are able to demonstrate that bursting events require many more active modes and therefore a dynamical cycle must support their nonlinear interactions. This serves as some substantiation for the use of triad variables in recurrent flow analysis. 
It should also be remembered that these variables encode a natural symmetry reduction, the amplitudes and triad phases being the ``dynamical degrees of freedom''. In a similar way to the method of slices \cite{Willis:2013bu} this removes the need to search over translations for relative solutions, thereby making the method significantly more computationally efficient. 

It is has also been demonstrated that synchronisation and alignment of triad phases can play a crucial role in dictating energy fluxes through Fourier modes in such systems \cite{Bustamante:2014jf, Buzzicotti:2016bz, murray_2018}. It therefore stands to reason that our recurrence function based on triad amplitudes is well-placed to diagnose near recurrence in situations when strong transfers of energy are occurring. 

A further indication of the value and originality of these recurrence results is the observation that a large number of the recurrences identified with $R_\xi,$ including those corresponding to bursting events, have values of $R_\phi$ far in excess of the threshold (i.e. $R_\phi \sim 1$). In other words, as indicated in figure \ref{fig:R_t_tau}, there is no evidence of a near recurrence occurring  in the classical recurrence function, and no amount of processing of guesses, or tuning of thresholds, will find them. Our results  disproves the hypothesis that these flows, being highly unstable, are not visited closely enough and for long enough for a recurrence to be identified. We do acknowledge that such problems will likely be encountered at higher Reynolds numbers and that other methods will probably be necessary, however, the results shown here demonstrate that careful variable choice can be critical in any method seeking to identify recurrent flows. 

There is some additional overhead in computing a relevant set of triads in a given system. For high resolutions computing the set of triads can be a computationally expensive exercise, but it is worth noting that this pre-processing is only required once for each resolution. In this work we were able to use a reduced set of linearly independent triads with good success. We also note that any set of triads, whether it is ``full'' and forms a linearly independent basis or not, will not be unique. Further work is needed to determine how sensitive these results are to the protocol used to construct the cluster matrix, and how such choices are developed for larger systems (both as $Re$ increases and in larger geometries).

While the computation is involved, the resulting data for the set of triads that we have used is quite small, ($655\times 6 = 3930$ integers) so it can  easily be shared (see supplementary material). One priority for future research in this area is to collate a repository of triad sets which have shown value in such studies.

The second major result of this paper is that, armed with an appropriately distributed set of recurrent flows, statistics of turbulence are very well predicted from their weighted average. In particular we have shown that fewer orbits than expected are necessary to obtain a reasonable reconstruction, and that pruning out ``unnecessary'' orbits can make correlations between individual weights and the instability of the states clearer. We suggest that weights proportional to the reciprocal of the sum of the real parts of unstable Floquet exponents provide as good a reconstruction of the statistics as those computed by minimising KL-divergence. Remaining inaccuracy in the reconstruction, e.g. in the middle of the dissipation and energy input distribution or the tails of the energy distributions, is more than likely due to missing one or two key recurrent flows in that region. We have confirmed this by conducting a short recurrent flow analysis (over $10^4$ time units) focusing on shorter period orbits ($T<25$) during which we found five new orbits occupying the interval $0.15<D/D_{lam}<0.25$ (not shown).  We might also conjecture that a more optimal set of triads, or using different sets to target particular parts of the turbulent cascade, would improve the coverage further. Our results suggest that targeting of solutions will confer a significant efficiency saving when only a handful of the best UPOs can serve to recreate the turbulent statistics.



\noindent
{\em Acknowledgements}. 

DL was supported by EPSRC New Investigator Award EP/S037055/1 ``Stabilisation of exact coherent structures in fluid turbulence'', ER was supported by a Keele University PGR studentship, AL was supported by the St Andrews Research Internship Scheme. Special thanks to Dr Miguel Bustamante for helpful discussions.

\vspace{1cm}

 The authors report no conflict of interest.
 
\bibliography{papers}

\end{document}